\documentclass{aastex62}
\usepackage{amsmath}
\usepackage{color}

\received{2019 February 18}
\revised{2019 August 7}
\accepted{2019 August 26}
\published{2019 September 27}
\submitjournal{ApJ}
\shorttitle{X-ray mechanism on kiloparsec-scales in PKS~1127$-$145}
\shortauthors{Butuzova and Pushkarev}

\begin{document}

\title{Inverse Compton Scattering of the Central Source Photons \\ as an X-ray Emission Mechanism on Kiloparsec Scales in PKS~1127$-$145}

\correspondingauthor{Marina S. Butuzova}
\email{mbutuzova@craocrimea.ru}

\author[0000-0001-7307-2193]{Marina S. Butuzova}
\affil{Crimean Astrophysical Observatory, Nauchny 298409, Crimea, Russia}

\author[0000-0002-9702-2307]{Alexander B. Pushkarev}
\affiliation{Crimean Astrophysical Observatory, Nauchny 298409, Crimea, Russia}
\affiliation{Astro Space Center of Lebedev Physical Institute,
Profsoyuznaya 84/32, Moscow 117997, Russia}

\begin{abstract}

The beamed inverse Compton/cosmic microwave background model has generally been used for the interpretation of X-ray radiation from kiloparsec-scale jets of the core-dominated quasars.
Recent \textit{Fermi}-LAT and \textit{HST} observations have brought this model into question.
We examine the assumption that X-rays from the kiloparsec-scale jet of the quasar PKS~1127$-$145 are produced by inverse Compton scattering of the central source emission.
In this context, we show that both similarity and distinction between the observed radio and X-ray spectral indices for some of the jet knots can be explained under a single power-law electron energy distribution.
We derive that the viewing angle of the kiloparsec-scale jet is about $35^\circ$ and the jet has a moderate relativistic speed of $\approx 0.8c$.
The predicted gamma-ray flux of the jet is found to be a few orders of magnitude lower than the minimum flux level measured by \textit{Fermi}-LAT, further supporting our scenario.
\end{abstract}

\keywords{galaxies: quasars --- individual --- PKS~1127$-$145, relativistic jets --- radiation mechanisms: non-thermal
}

\section{Introduction} \label{sec:intro}

Kiloparsec-scale jets of active galactic nuclei (AGNs) have been observed in X-rays by the \textit{Chandra X-ray Observatory} since 2000 \citep{Schwartz2000}. 
An angular resolution of $\sim0.5''$ in the X-ray range corresponds to that in the radio (e.g., MERLIN, VLA) and optical bands.
This allowed us to draw conclusions about X-ray emission mechanisms by analyzing multiwavelength data \citep[see][for a review]{Harris06}.
The first quasar detected by \textit{Chandra}, PKS~0637$-$752, showed an X-ray flux significantly greater than that expected from extrapolation from the radio-to-optical synchrotron spectrum to high frequencies. 
This fact excluded a hypothesis on synchrotron radiation from electrons with a single power-law energy distribution, emitting in the radio-optical band, as an X-ray emission mechanism.
After analyzing other possibilities, \citet{Schwartz2000} concluded that most probably X-rays are produced by inverse Compton scattering (ICS) on self-synchrotron photons, and the magnetic field is far below than that required by the equipartition. 
However, the authors noted that the magnetic field distribution in the jet can be non-uniform and almost all radio emission can be generated in a small region with the high magnetic field, while X-rays come from a significantly larger area.     

\citet{Tav00} and \citet{Cel01} suggested an alternative scenario in which X-rays from the PKS~0637$-$752 jet are generated by ICS of cosmic microwave background photons (IC/CMB). 
Then, for a priori fulfillment of the equipartition, both a small viewing angle and ultra-relativistic velocity of the kiloparsec-scale jet are needed.
The main arguments in favor of this assumption were: (1) the apparent one-sidedness of the jet due to relativistic effects; (2) apparent superluminal motion detected for the parsec-scale jet of PKS~0637$-$752 \citep{Chartas00,Edw06}.   
This beamed IC/CMB model was widely applied for the interpretation of X-ray emission from the core-dominated quasar jets and  investigation of physical conditions therein \citep[see, e.g.,][]{GeorKaz04, KatSta05, Harris06, Marsh11} as well as for the selection of new targets for \textit{Chandra} observations \citep{Samb04, Marsh05, HoganList11, Marsh18}.

The beamed IC/CMB model was ruled out for 3C~273 \citep{MeyGeor14} and PKS~0637$-$752 \citep{Meyer15} because the predicted high level of steady gamma-ray flux has not been detected from \textit{Fermi}-LAT observations. 
Alternative X-ray emission mechanisms, such as synchrotron radiation from a second high-energy electron population \citep{DA04} or proton synchrotron radiation \citep{Ahar02},  require the simultaneous action of two mechanisms of particle acceleration, complicating the concept of a jet and adding a few free parameters. Variations of these parameters allow one to obtain reasonable estimates of jet physical conditions.
In this paper, we re-examine ICS of the central source radiation (IC/CS) as a possible X-ray emission mechanism for the kiloparsec-scale jet knots located at distances smaller than $\approx20''$ from the PKS~1127$-$145 nucleus. 
We adopt the $\Lambda$CDM-model with $H_0=71$~km~s$^{-1}$~Mpc$^{-1}$, $\Omega_{\text{m}}=0.27$, $\Omega_\Lambda=0.73$~\citep{Komatsu09}.

The layout of the paper is as follows. In Section~\ref{sec:iccspos} we list the factors that make IC/CS a possible X-ray emission mechanism of the quasar kiloparsec-scale jets. At the beginning of Section~\ref{sec:iccs}, we write the basic formulae for IC/CS and describe the used parameters. Then we derive the estimates of the physical parameters for two bright jet knots detected both in X-ray and at radio frequencies $\lesssim5$~GHz. The inclination angle and speed of the kiloparsec-scale jet are inferred in Section~\ref{sec:kpcangle}. Section~\ref{sec:gam} includes the prediction of gamma-ray flux for the jet under IC/CS. A discussion and conclusions are presented in Sections~\ref{sec:disc} and \ref{sec:conc}, respectively.

\section{Is Domination of IC/CS Over IC/CMB Possible?}
\label{sec:iccspos}

\begin{figure}
\begin{center}
\includegraphics[scale=0.7]{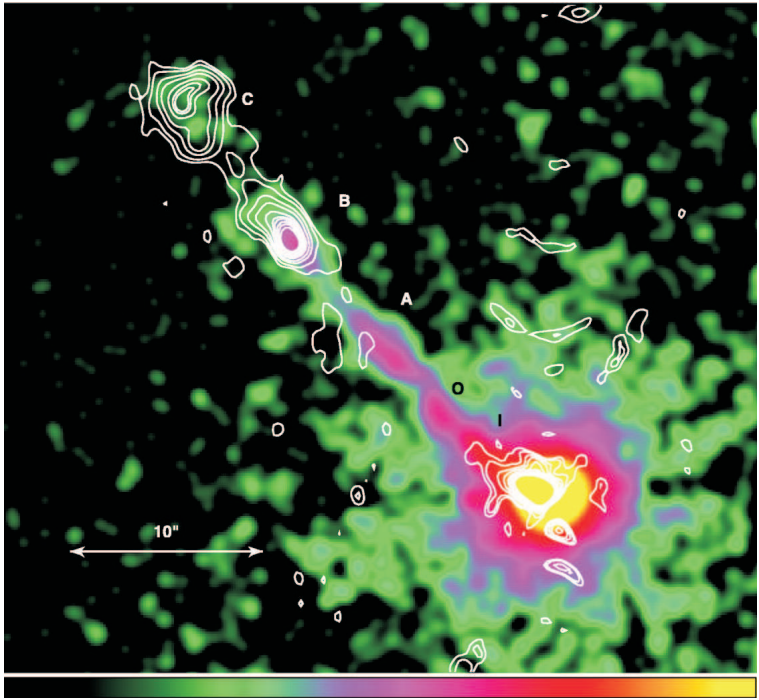}  
\caption{Distributions of X-rays (color) and radio emission at 8.5~GHz (contours) along a kiloparsec jet of the quasar PKS~1127$-$145 from~\citet{Siem07}. 
Color wedge ranges from $10^{-10}$ to $10^{-5}$~photons~cm$^{-2}$~s$^{-1}$~pixel$^{-2}$ indicating the X-ray brightness.
The radio contours are plotted at 0.2, 0.3, 0.5, 0.7, 1.0, 1.2, 1.3, 1.5, 1.7, and 2.0~mJy~beam$^{-1}$. Symbols I, O, A, B, C mark kiloparsec-scale jet knots.  }
\end{center}
\label{fig:kpcjet}
\end{figure}

The morphology of the kiloparsec-scale jet in quasar PKS~1127$-$145  ($z=1.184$, 8.33~kpc/arcsec) in the radio and X-ray bands was described in \citet{Siem02, Siem07}.
The X-ray intensity reaches its maximum at the first knot~I and decreases with distance from the core (Fig.~\ref{fig:kpcjet}). Hereinafter,  by the term ``knot" we mean a region of increased surface brightness on kiloparsec scales. At the terminal knot~C, the X-ray intensity is low and its distribution within the knot has no prominent maximum \citep{Siem07}. 
Because of the quadratic decline in energy density of the central source  with distance, the distribution of X-ray intensity along the jet can be explained within the assumption that in the inner knots X-rays are formed by ICS of central source photons, while IC/CMB possibly acts in the most distant knot~C. By the central source (CS) we mean the active nucleus, parsec-scale jet, and their surroundings. 

In the radio domain, knots~B and C are only detected at frequencies ranging from 1.4 to 8.5~GHz. Knot~A appears at 1.4 and 5~GHz, while knot~O manifests itself only at 1.4~GHz \citep{Siem07}. It can be interpreted by electron energy losses caused by IC/CS increase at progressively smaller distances from the CS. This results in a steepening spectrum of higher energy electrons \citep{Pachol}. Since the energy of electrons radiating within the \textit{Chandra} frequency range by IC/CS is much lower than that of electrons producing synchrotron radio emission, the relation $\alpha_{\text{R}}= \alpha_{\text{X}}+0.5$ will be expected. This is true for observed radio $\alpha_{\text{R}}$ and X-ray $\alpha_{\text{X}}$  spectral indices of knot~A \citep{Siem07} as evident from Table~\ref{tab:input}. Knot~I is distinctly detected at 8.5~GHz, while at lower frequencies the angular resolution is not enough to separate it from the bright quasar core.
We constrain physical parameters of knots~A and B only because these are detected and have known spectral indices in both radio and X-ray bands \citep{Siem07}.

\begin{figure}
\begin{center}
\includegraphics[scale=0.6]{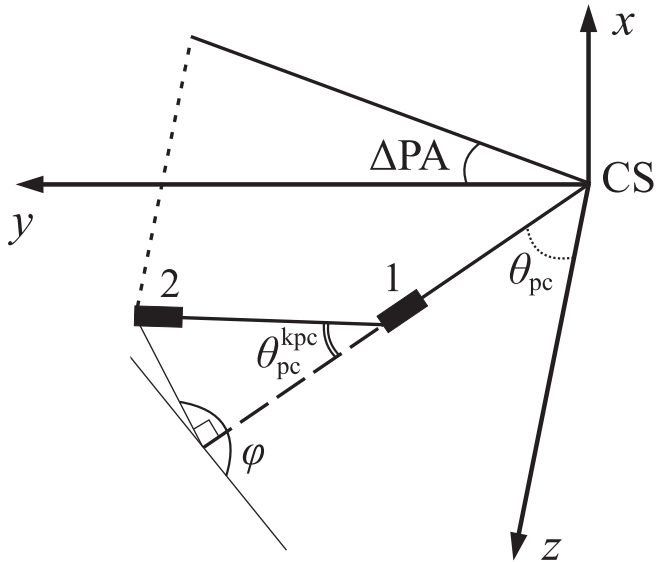}  
\caption{ 
Schematic of the jet bend between the parsec and kiloparsec scales in the observer's frame. $\theta_{\text{pc}}$ is the parsec-scale jet angle to the line of sight, $\theta_{\text{pc}}^{\text{kpc}}$ is the initial angle of the jet bend, $\varphi$ is the azimuth angle of the bend, and ``1'' and ``2'' denote parsec- and kiloparsec-scale jets, respectively.
Axes $x$ and $y$ lie in the plane of the sky. Axis $z$ is directed along the line of sight.
Directions of jets on the parsec and kiloparsec scales correspond to their observed position angles. $\Delta \text{PA}$ is the difference of these angles. The figure illustrates an initial jet bend between parsec and kiloparsec scales. If the speed of the kiloparsec-scale jet is lower than that of the parsec-scale jet, the angle between the kiloparsec-scale jet and the line of sight in the observer's frame increases due to relativistic aberration.  
 }
\end{center}
\label{fig:shbend}
\end{figure}

For radio galaxies, ICS of CS photons on jet or lobe electrons has been widely discussed as a possible mechanism for generation of X-rays from kiloparsec-scale jets \citep{Cel01,StawOB08,Ostor10,Migliori12}.
Considering X-ray emission mechanisms of quasar kiloparsec-scale jets, ICS of CS is usually excluded, as the jet viewing angle  is believed to be $\lesssim10^\circ$ \citep[see, e.g.,][and references therein]{Harris06}. This small value of the angle is required to explain an apparent absence of the counter-jet by Doppler de-boosting, assuming equality of the jet and counter-jet. Then, the small kiloparsec-scale jet viewing angle results in a large de-projected distance of the jet from the CS.
Due to a quadratic decline of the total energy density of CS $W_{\text{CS}}$ with distance to CS, in kiloparsec-scale jets $W_{\text{CS}}$  is smaller than the energy density of the CMB, $W_{\text{CMB}}$. 
But the observed radiation of active nuclei from radio to millimeter wavelengths is mostly generated in their parsec-scale jets \citep{Kovalev2005}.
The emission flux density of a parsec-scale jet is subject to Doppler boosting for an Earth's observer because the jet moves with ultra-relativistic speed at a small viewing angle $\theta_{\text{pc}}$ 
\citep[see e.g.,][]{List16}.
Jets have a bend between parsec and kiloparsec scales or, as for PKS~1127$-$145, a bend on the kiloparsec scales also \citep{Siem02}.
The bend angle has an average value of $\theta_{\text{pc}} / 1.4$, derived from a sample of 50 core-dominated quasars~\citep{But16}.
Interpreting the observed difference between the position angle\footnote{Position angle (PA) is the angle between directions from the core to north and to the jet feature measured toward the rise of R.A.} of the parsec-scale jet \citep{List16} and the kiloparsec-scale jet end \citep{Siem02} as an outflow bend (see Fig.~\ref{fig:shbend}), and using the formula (1) in \citet{CM93}\footnote{\cite{CM93} denote the difference in position angles as $\eta$, $\theta_{\text{pc}}^{\text{kpc}}$ as $\zeta$, $\theta_{\text{pc}}$ as $\theta$.}, we obtained that for most values of the azimuth angle $\varphi$ the initial jet bend is in a narrow range of $2^\circ-3^\circ$. 
Despite the fact that the bend is present, the angle of the initial jet bend $\theta_{\text{pc}}^{\text{kpc}}$ is less than $\theta_{\text{pc}}$.
Therefore, Doppler boosting of the parsec-scale jet emission for an observer in the kiloparsec-scale jet is larger than for an Earth observer. This can lead to $W_{\text{CS}}>W_{\text{CMB}}$.

Second, the scattering of photons with frequency of the CMB spectral maximum provides the main contribution to the scattered emission under IC/CMB. In the case of IC/CS, the spectrum of scattering radio to millimeter photons is a power-law or, as for PKS~1127$-$145, it can be approximated by several power-law parts.
For the scattering of CS emission on electrons that have a power-law energy spectrum, energies of interacting particles, which give the main contribution to the observed X-rays, are defined by the spectral indices of photon and electron energy distributions (see Sec.~\ref{sec:iccs}).
The frequency of these CS photons is not coincident with that of the CMB maximum. Hence, electrons that upscatter photons of CS and CMB to the observed frequency have different energy, which can create conditions for domination of IC/CS over IC/CMB.
Therefore, to define a type of scattering photon, in our view it is more accurate to compare a spectral flux of the scattered radiation expected in IC/CMB and IC/CS.
In the case of IC/CMB, the flux density of the observed scattered emission $F_{\text{CMB}}$ can be obtained through electron energy losses for radiation \citep[e.g.][]{MBK10}. 
However, in the case of IC/CS we should take into account the spectrum of soft photons and the facts that (i) their momenta are distributed within a small solid angle and (ii) particle density in the jet knots depends on their distance from the CS.
Thus, to find an expression for the IC/CS flux density, we used the invariant Boltzmann transport equation for ICS (see Appendix~\ref{sec:app}).

\section{Inverse Compton Scattering of the Central Source Photons} \label{sec:iccs}

Consider ICS of the CS emission on the kiloparsec-scale jet electrons.
The scattering photon frequency distribution can be found from the observed spectral flux density of the quasar, $F(\omega)$.
As the CS radiation from the radio to millimeter wavelengths mainly originates on parsec scales \citep{Kovalev2005}, the spectral luminosity of the CS in the parsec-scale jet frame is $L_\omega=4 \pi D_L^2 (1+z)^{3+a} \delta^{-3-\alpha}F(\omega)$, where $D_L$ is the luminosity distance, $\alpha$ is the spectral index of radiation under its power-law approximation,  $F(\omega)=Q\omega^{-\alpha}$, $\delta=\sqrt{1-\beta_{\text{pc}}^2}/(1-\beta_{\text{pc}}\cos \theta_{\text{pc}})$ is the Doppler factor for an Earth observer, and $\beta_{\text{pc}}$ is the bulk parsec-scale jet velocity in units of the speed of light $c$. 
On the other hand, $L_\omega=4 \pi (R/\sin \theta)^2 \delta_{\text{j}}^{-3} F^{\text{j}}(\omega_{\text{j}})$, where $R$ is the distance from the CS to some kiloparsec-scale jet knot projected on the sky, $\theta$ is the kiloparsec-scale jet viewing angle, $\delta_{\text{j}}$ and $F{^{\text{j}}}(\omega_{\text{j}})$ are the Doppler factor and the CS flux observed in the kiloparsec-scale jet, respectively, and
\begin{equation}
 \omega_{\text{j}}=\omega(1+z)\delta_{\text{j}}/\delta 
\label{eq:wj}
\end{equation} 
is the scattering photon frequency in the kiloparsec-scale jet frame.
Comparing expressions for $L_\omega$ and inserting $F^{\text{j}}(\omega)=c \hbar \omega_{\text{j}} N(\omega_{\text{j}})$, we obtain the distribution of scattering photons per unit of solid angle
\begin{equation}
N(\omega_{\text{j}})=\left((1+z)\frac{\delta_{\text{j}}}{\delta} \right)^{3+\alpha} \frac{D_L^2 \sin^2 \theta}{c \hbar R^2} Q \omega_{\text{j}}^{-1-\alpha}.
\label{eq:Nwj}
\end{equation}
For the case of PKS~1127$-$145, the measured apparent velocity $\beta_{\text{app}}\approx 12$ \citep{List16} corresponds to $\theta_{\text{pc}}\sim1/\Gamma_{\text{pc}}\approx(1+\beta_{\text{app}}^2)^{-0.5}\approx5^\circ$, $\beta_{\text{pc}}\approx0.995$ and $\delta\approx11$. 
To calculate $\delta_{\text{j}}$, one needs to use the angle between parsecc- and kiloparsec-scale jets in the reference frame of the kiloparsec-scale jet, $\theta^{\prime\, \text{kpc}}_{\,\,\,\text{pc}}$. But the kiloparsec-scale jet speed is unknown. Based on investigations of \cite{WA97, ArLon04, MullinH09, Homan15}, we assume a moderately relativistic speed of the kiloparsec-scale jet. Hence, for simple estimation of its parameters we used  $\theta_{\text{pc}}^{\text{kpc}}$ as the angle of initial jet bend between parsec and kiloparsec scales (i.e., without taking into account increase of the kiloparsec-scale jet viewing angle in the plane containing the line of sight due to the jet deceleration). In Section \ref{sec:kpcangle} we describe how the kiloparsec-scale jet parameters change if this angle is  three times larger.
To estimate $\theta_{\text{pc}}^{\text{kpc}}$ we followed Equation (1) in \citet{CM93} for the difference between the parsec-scale jet \citep{List16} and the part of kiloparsec-scale jet located at 2$''$ from the CS \citep{Siem02}. Changing the azimuth angle (see Fig.~\ref{fig:shbend}) from $0^\circ$ to $360^\circ$, we obtained that for most cases the jet bend is about $1^\circ$. For further calculations we use $\theta_{\text{pc}}^{\text{kpc}}= 1^\circ$ as a lower limit and, as we noted in Section~\ref{sec:iccspos}, it is unlikely that this angle exceeds $3^\circ$.

\begin{figure}
\begin{center}
\includegraphics[scale=0.6]{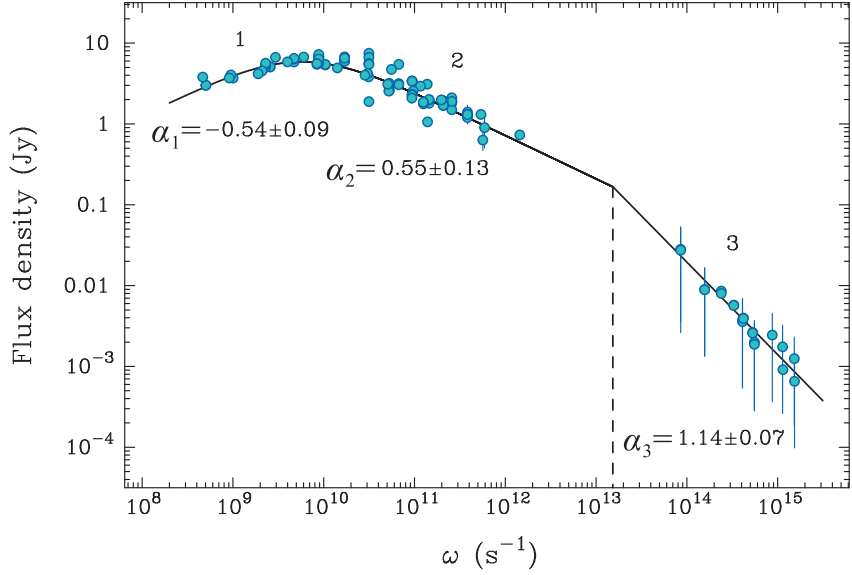}
\caption{Observed central source spectrum fitted by three power-law parts represented by solid line. Parameters of the fits are listed in \autoref{tab:input}.
The dashed line indicates the frequency of spectral break $\omega_{\text{br}}$. }  
\end{center}
\label{fig:CS_spec}
\end{figure}

Figure~\ref{fig:CS_spec} is based on data from \cite{HealeyR07, Dixon70, KuehrW81, Condon98, WrightG91,MurphyS10, Ade14, Wright09}. It shows that the radio-optical spectrum of the quasar PKS~1127$-$145 has its maximum at frequency $\omega_0\approx 5.65\cdot10^{10}$~s$^{-1}$ and a break at a frequency for which there are no observational data. 
In intervals between the spectral maximum and the break we approximate the spectrum by a power-law with fitted parameters listed in Table~\ref{tab:input}.
We define parts~1 and 2 as positioned before and after $\omega_0$, respectively, and part~3 as positioned beyond the spectral break (see Fig.~\ref{fig:CS_spec}).
The frequency of the break $\omega_{\text{br}}\approx 1.87\cdot 10^{13}$~s$^{-1}$ was found as a crossing point of the two power-law parts and assuming that the CS spectrum is continuous.
Then, we obtain (see \autoref{sec:app})
\begin{equation}
F(\omega')=(1+z)^{3+\alpha}\left(\frac{\delta_{\text{j}}}{\delta}\right)^{3+\alpha}\frac{r_e^2}{2}V\frac{\sin ^2 \theta}{R^2} \left(m_e c^2 \right)^{1-\gamma} \mathcal{K} Q \omega' \iint { \Gamma^{-\gamma-2} \omega_{\text{j}}^{-\alpha-2} d\Gamma d\omega_{\text{j}}},
\label{eq:Fcs_int}
\end{equation}
where $\omega'$ is the scattered photon frequency without a correction for redshift, $V=x y^2/\sin \theta$ is the knot volume, $x$ and $y$ are the longitudinal and transverse projected sizes, respectively (see Table~\ref{tab:input}), 
$\gamma$ and $\mathcal{K}$ are the spectral index and the constant of power-law knot electron energy distribution, and $\Gamma=E/(m_e c^2)$ is the electron's Lorentz factor.   
As a result of integration of power-law functions in Equation~(\ref{eq:Fcs_int}) one of limits is important. It is determined by the relation between spectral indices of scattering radiation $\alpha$ and electron energy distribution $\gamma$. 
For integration limits, we use either physical constraints of photon and electron power-law spectra, or the limits
\begin{eqnarray}
\Gamma^2&\ge \frac{\omega'}{2 \omega_{\text{j}} (1-\beta \cos \theta)},\nonumber \\
\omega_{\text{j}}&\ge \frac{\omega'}{2 \Gamma^2 (1-\beta \cos \theta)},
\label{eq:limit_int}
\end{eqnarray}
resulting from integrating Equation~(\ref{eq:kineq2}) using the Dirac delta function.

\begin{table}
\caption{Observed Parameters of the Central Source and Jet Knots of the Quasar PKS~1127$-$145.
The CS spectrum is approximated by the power law $F=Q\omega^{-\alpha}$ within the three specified frequency ranges. The bottom part of the table contains only the knots that we analyze in this paper. We specify a 5~GHz flux density for those knots that we use in calculations.}
\label{tab:input}
\renewcommand{\arraystretch}{1.1}
\begin{center}
\begin{tabular}{cccccccc}
\hline
\multicolumn{8}{c}{Central source emission}\\
\hline
Part & \multicolumn{3}{c}{Frequency range,$\omega$,  s$^{-1}$}& \multicolumn{2}{c}{$Q$, erg~cm$^{-2}$~s$^\alpha$ } & \multicolumn{2}{c}{Spectral index $\alpha$} \\
\hline
1 & \multicolumn{3}{c}{$<5.65\cdot10^{10}$} & \multicolumn{2}{c}{$5.93\cdot10^{-28}$} & \multicolumn{2}{c}{$-0.54\pm 0.09$} \\
2 & \multicolumn{3}{c}{$5.65\cdot10^{10}-1.87\cdot10^{13}$} & \multicolumn{2}{c}{$2.89\cdot10^{-17}$} & \multicolumn{2}{c}{$0.55\pm 0.13$} \\
3 & \multicolumn{3}{c}{$>1.87\cdot 10^{13}$} & \multicolumn{2}{c}{$1.95\cdot10^{-9}$} & \multicolumn{2}{c}{$1.14\pm 0.07$}\\
\hline
\multicolumn{8}{c}{Kiloparsec-scale jet\footnote{Observed data from \citet{Siem02,Siem07}}} \\
\hline
\multicolumn{2}{c}{~} & \multicolumn{2}{c}{Knot A} & \multicolumn{2}{c}{Knot B} & \multicolumn{2}{c}{Knot C} \\
\hline
\multicolumn{2}{l}{Projected distance from CS} & \multicolumn{2}{c}{$11.2''$} & \multicolumn{2}{c}{$18.6''$} & \multicolumn{2}{c}{$28.5''$} \\
\multicolumn{2}{l}{Flux density at 5 GHz (mJy)} & \multicolumn{2}{c}{$1.2\pm0.2$} & \multicolumn{2}{c}{$14.4\pm1.4$} & \multicolumn{2}{c}{$16.7\pm1.7$} \\
\multicolumn{2}{l}{2 keV flux (nJy)} & \multicolumn{2}{c}{$0.23\pm0.1$} & \multicolumn{2}{c}{$0.11\pm0.11$} & \multicolumn{2}{c}{$0.022\pm0.068$}\\
\multicolumn{2}{l}{Radio spectral index $\alpha_{\text{R}}$} & \multicolumn{2}{c}{$1.32\pm0.17$} & \multicolumn{2}{c}{$0.91\pm0.07$} & \multicolumn{2}{c}{$0.85\pm0.08$} \\
\multicolumn{2}{l}{X-ray spectral index $\alpha_{\text{X}}$} & \multicolumn{2}{c}{$0.66\pm0.15$} & \multicolumn{2}{c}{$1.0\pm0.2$} & \multicolumn{2}{c}{$1.2\pm0.6$} \\
\multicolumn{2}{l}{Size $(x''\times y'')$} & \multicolumn{2}{c}{$2.6''\times0.7''$} & \multicolumn{2}{c}{$2.8''\times0.9''$} & \multicolumn{2}{c}{$2.4''\times0.6''$}\\
\hline
\end{tabular}
\end{center}
\end{table}

\subsection{Physical Parameters of Knot B}  \label{subsec:phpB}

We begin our consideration with knot~B, showing equal spectral indices of radio and X-ray emission.
To determine the physical parameters of knot~B from Equation~(\ref{eq:Fcs_int}), we make some assumptions on the upper, $\Gamma_{\text{max}}$, and lower, $\Gamma_{\text{min}}$, boundaries of a power-law electron energy distribution.
For this purpose, we use the relation between energies of interacting particles under ICS \citep{Pachol}:
\begin{equation}
\Gamma^2=3(1+z)\omega_{\text{X}}/(4\omega_{\text{j}}). 
\label{eq:osv}
\end{equation}

\begin{figure}
\begin{center}
\includegraphics[scale=0.63]{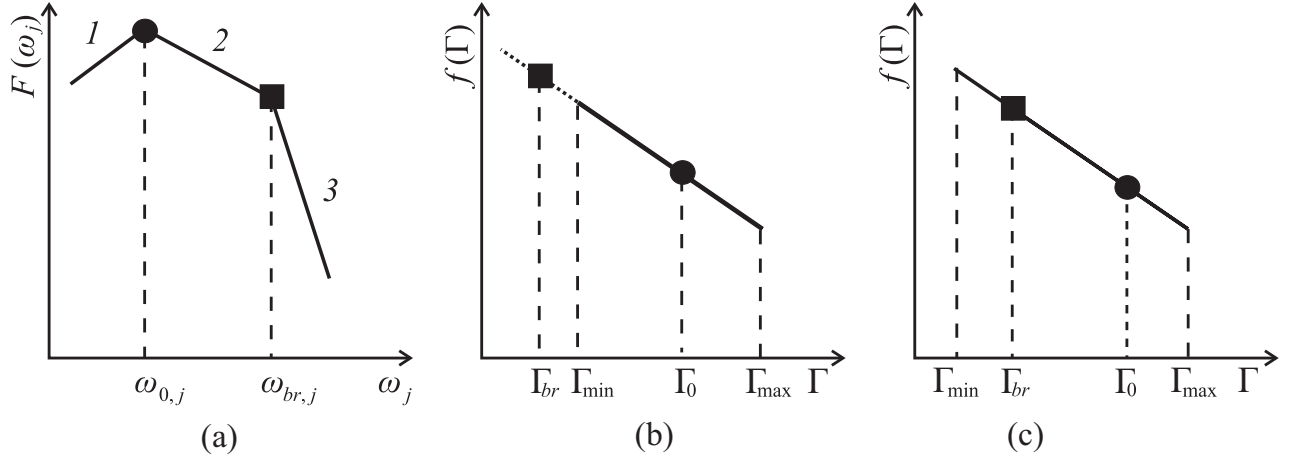}
\caption{Schematics of the scattering photon spectrum (a) and two possible electron distributions in knot~B.
Case (b) corresponds to the restriction by electron distribution for IC/CS.
This situation is true for knot~A.
Case (c) illustrates the restriction by the photon spectrum.
The filled circles denote the frequency of spectral maximum (a) and the Lorentz factor of electrons scattering these photons to the fixed X-ray frequency $\omega_{\text{X,\,obs}}$ (b, c). The filled squares mark the photons with the frequency of the spectral break (a) and the Lorentz factor of electrons scattering these photons to $\omega_{\text{X,\,obs}}$.}
\label{fig:ph_e_distr}
\end{center}
\end{figure}

First, we suppose $\Gamma_{\text{max}}>\Gamma_0$ and $\Gamma_{\text{min}}<\Gamma_0$ (Fig.~\ref{fig:ph_e_distr}).
Electrons with $\Gamma_0$ Lorentz factor change the photon frequency $\omega_{0,{\text{j}}}$ to the observed X-ray frequency $\omega_{\text{X}}=3\cdot 10^{18}$~s$^{-1}$ (2~keV) by ICS.
Then, under ICS of part~1 of the CS spectrum, photons having the upper boundary frequency $\omega_{0,{\text{j}}}$ are scattered by electrons with a Lorentz factor differed from boundary values, i.e., not all electrons scatter photons of the part~1 of the CS spectrum at the frequency $\omega_{\text{X}}$.
That is, the energy of interacting particles is restricted by the photon spectrum.
Therefore, we first integrate Equation~(\ref{eq:Fcs_int}) over $d\Gamma$. In this case, the lower limit given by Equation~(\ref{eq:limit_int}) dominates. 
Because $\gamma-2\alpha_1-1>0$, the redshift-corrected X-ray flux in the case of scattering of part~1 of the CS spectrum is 
\begin{equation}
F_1 (\omega_{\text{X}})=(1+z)^{\alpha_1-(\gamma-1)/2} \left(\frac{\delta_{\text{j}}}{\delta} \right)^{3+\alpha_1} r_e^2 \frac{V \sin^2 \theta}{R_{\text{B}}^2} (m_e c^2)^{1-\gamma} \mathcal{K} Q_1 \frac{\left[ 2 (1-\cos \theta)\right]^{(\gamma+1)/2}}{(\gamma+1)(\gamma-2 \alpha_1-1)} \omega_{0,{\text{j}}}^{(\gamma-1)/2-\alpha_1} \omega_{\text{X}}^{-(\gamma-1)/2},
\label{eq:F1B}
\end{equation}
where $R_{\text{B}}=18.6''=4.76 \cdot 10^{23}$~cm is the  projected distance of knot~B from the CS (hereinafter literal indices of parameters indicate a knot under consideration).
From Equation~(\ref{eq:F1B}) it follows that \textit{if there is a restriction by the photon spectrum, the X-ray spectral index is equal to the radio spectral index} $\alpha_{\text{X}}=\alpha_{\text{R}}=(\gamma-1)/2$.
In this case, the scattering of photons with a given frequency provides the most relevant contribution to the scattered flux. It corresponds to the case of scattering of the monochromatic radiation field on electrons having the power-law energy spectrum.

The second assumption about the boundary of the electron energy spectrum is that $\Gamma_{\text{min}}$ is higher than the Lorentz factor $\Gamma_{\text{br}}\approx250$ needed to change the photon frequency from $\omega_{\text{br, j}}$ to $\omega_{\text{X}}$ (see Fig.~\ref{fig:ph_e_distr}b).
Then, not all photons corresponding to part~2 of the CS spectrum are scattered by electrons at frequency $\omega_{\text{X}}$.
That is, the energy of interacting particles is restricted by the electron spectrum.
Therefore, we first integrate Equation~(\ref{eq:Fcs_int}) over $d\,\omega$. 
As $2\alpha_2-\gamma_{\text{B}}+1<0$, 
the flux is
\begin{equation}
F_2(\omega_{\text{X}})=\left(\frac{\delta_{\text{j}}}{\delta} \right)^{3+\alpha_2}\frac{r_e^2}{2}\frac{V \sin^2 \theta}{R_{\text{B}}^2} (m_e c^2)^{1-\gamma} \mathcal{K} Q_2 \frac{\left[2 (1-\cos \theta) \right]^{\alpha_2+1}}{(\alpha_2+1)(\gamma-2 \alpha_2-1)} \Gamma_{\text{min}}^{2 \alpha_2-\gamma+1} \omega_{\text{X}}^{-\alpha_2}.
\label{eq:F2eB}
\end{equation}
Equation~(\ref{eq:F2eB}) shows that \textit{if there is a restriction by the electron spectrum, the X-ray spectral index is equal to the spectral index of scattering emission} $\alpha_{\text{X}}=\alpha_2$.
As the scattering by electrons with fixed energy makes the main contribution to the scattered flux, this case corresponds to the scattering of the power-law photon spectrum on monoenergetic electron distribution. 
To reconcile the observed X-ray spectrum with $\alpha_{\text{X}}\approx\alpha_{\text{R}}\approx0.9$ \citep{Siem07}, the condition $F_1/F_2\gg1$ should be fulfilled.
This is possible if $\Gamma_{\text{min}}>3\cdot10^5$, which is implausible.

Therefore, $\Gamma_{\text{min}}<\Gamma_{\text{br}}$ (Fig.~\ref{fig:ph_e_distr}c) and for ICS of parts~2 and~3 of the CS spectrum there is a restriction by the photon spectrum.
Substituting the corresponding values $Q$ and $\alpha$ (see Table~\ref{tab:input}) into Equation~(\ref{eq:Fcs_int}) we found that for ICS of both these parts, the scattering of photons with the same frequency $\omega_{\text{br, j}}$ gives the main contribution to the scattered fluxes $F_2$ and $F_3$. 
As follows from Eq.~(\ref{eq:osv}), electrons with the same energy scatter photons with fixed frequency to the observed X-ray frequency. Hence, the fluxes $F_2$ and $F_3$ are expected to be equal and either of them can be used as the flux produced by ICS of parts~1 and 2 of the CS spectrum.
The flux $F_1$ was neglected because it is about $1\%$ of $F_2$ or $F_3$.
Then, the expected X-ray flux from knot~B under IC/CS is
\begin{equation}
F_{\text{X}}=(1+z)^{\alpha_3-(\gamma-1)/2} \left(\frac{\delta_{\text{j}}}{\delta} \right)^{3+\alpha_3} r_e^2 \frac{V \sin^2 \theta}{R_{\text{B}}^2} (m_e c^2)^{1-\gamma} \mathcal{K} Q_3 \frac{\left[ 2 (1-\cos \theta)\right]^{(\gamma+1)/2}}{(\gamma+1)(2 \alpha_3-\gamma+1)} \omega_{\text{br, j}}^{(\gamma-1)/2-\alpha_3} \omega_{\text{X}}^{-(\gamma-1)/2}.
\label{eq:FxB}
\end{equation}
Using in Equation~(\ref{eq:FxB}) the observed data~\citep{Siem07} (see Table~\ref{tab:input}) and kiloparsec-scale jet viewing angle  $\theta=35^\circ$~(see Sec.~\ref{sec:kpcangle}), we found the electron distribution constant $\mathcal{K}=7\cdot10^{-11}$~(in CGS units). 
This allows us to estimate: (i) the density of emitting electrons $n_e=\int\mathcal{K}E^{-\gamma}dE=\mathcal{K}/(\gamma-1)(m_e c^2)^{1-\gamma} \Gamma_{\text{min}}^{1-\gamma}\approx1\cdot10^{-3}$~cm$^{-3}$ for $\Gamma_{\text{min}}=100$; (ii) the magnetic field strength $B_\perp=2.2\cdot10^{-5}$~G based on the observed flux at the defined radio frequency by Formula~(9) in \citet{MBK10} (see Table~\ref{tab:output}).
For the X-rays originated by IC/CMB, $\mathcal{K}$ is found from Equation~(\ref{eq:Fcmb}) in the Appendix and as a result it gives roughly the same magnetic field $B_{\perp,{\text{CMB}}}=1.4\cdot10^{-5}$~G and electron density $n_{e,{\text{ CMB}}}=1.6\cdot10^{-3}$~cm$^{-3}$ (see Equations~(10) and (8) in \citet{MBK10}).

\begin{table}
\caption{Jet Knot Parameters in the Framework of IC/CS, Assuming a Jet Angle with Line of Sight $\theta=35^\circ$.}
\label{tab:output}
\renewcommand{\arraystretch}{1.1} 
\renewcommand{\tabcolsep}{0.3cm}
\begin{center}
\begin{tabular}{l|c|c|c|c}
\hline
~ & \multicolumn{2}{c}{Knot A} & \multicolumn{2}{|c}{Knot B}\\ 
\hline
$\Gamma_{\text{min}}$ & \multicolumn{2}{c}{$2.7\cdot10^2-1.5\cdot10^4$} & \multicolumn{2}{|c}{$<2.7\cdot10^2$}\\ 
$\Gamma_{\text{min}}$ used for estimates & $5\cdot10^2$ & $10^3$ & 10 & $10^2$ \\ 
$\mathcal{K}$, CGS units & $4.1\cdot10^{-13}$ & $1.2\cdot10^{-12}$ & \multicolumn{2}{c}{$7.0\cdot10^{-11}$} \\
$n_e$, cm$^{-3}$ & $1.4\cdot10^{-4}$ & $6.4\cdot10^{-5}$ & $6.9\cdot 10^{-2}$ & $1.0 \cdot 10^{-3}$ \\
$B$, $10^{-5}$ G & 2.0 & 1.3 & \multicolumn{2}{c}{2.2} \\
$B_{\text{eq}}$, $10^{-5}$ G & 21 & 16 & 7.2 & 4.4\\
\hline
\end{tabular}
\end{center}
\end{table}

Since the observed radio spectrum of knot~B is a power law, it is impossible to determine lower and upper frequencies of the synchrotron spectrum cutoff, which are traditionally used to estimate the equipartition magnetic field. 
Therefore, to make minimum assumptions on the physical parameters of knot~B, following  \citet{Pachol} we expressed whole electron energy in the emitting region, $E_e$, and the total synchrotron luminosity in terms of boundary Lorentz factors of electron energy distribution.
For this, we used the fact that the frequency of the synchrotron spectrum maximum for a single electron is $\nu=0.29 (3 e B)/(2 m_e c)\Gamma^2$, where $e$ is the electron charge.
Then, from equation $E_e=B^2 V\phi/(8 \pi)$ ($\phi$ is a part of the whole volume occupied by the magnetic field, here $\phi=1$) we obtain
\begin{equation}
B_{\text{eq}}=\left[ \frac{(3-\gamma)}{(\gamma-2)(1-\alpha_{\text{R}})} \frac{32 \pi^2 D_L^2 Q_{\text{R}}}{c_2 m_e c^2 V} \tilde{c}_1 \left(\Gamma_{\text{min}}^{-\gamma+2}-\Gamma_{\text{max}}^{-\gamma+2} \right) \right]^{1/(3+\alpha_{\text{R}})}(\sin \vartheta)^{-2/(3+\alpha_{\text{R}})},
\label{eq:Heq}
\end{equation}
where constants in CGS units are $c_2=2.37\cdot10^3$, $\tilde{c}_1=0.29(3e)/(4\pi m c)=1.22\cdot10^6$, $\vartheta$ is the angle between the magnetic field and the line of sight, and $Q_{\text{R}}$ and $\alpha_{\text{R}}$ are the proportionality coefficient and the spectral index of synchrotron emission of knot~B, respectively. 
Since $-\gamma+2<0$, the choice of $\Gamma_{\text{max}}$ value does not significantly influence the value of $B_{\text{eq}}$.
Neglecting the right multiplier in Equation~(\ref{eq:Heq}), which is about 3 for the chaotic magnetic field, we found $B_{\text{eq}}\approx 4.4\cdot10^{-5}$~G for $\Gamma_{\text{min}}=100$, which agrees with the value obtained above from the comparison of radio and X-ray fluxes.
To match the magnetic field found from the ratio of the radio to X-ray fluxes with its equipartition value, \citet{Siem02,Siem07} introduced a moderate relativistic bulk motion for the kiloparsec-scale jet.
But there is no need to fulfill the energy equipartition condition in light of new observational results obtained by the Earth-space radio interferometer \textit{RadioAstron}, which show extremely high brightness temperatures of BL~Lac \citep{Gomez16}, 3C~273 \citep{Kovalev16}, B0529+483 \citep{Pilipenko18}, AO~0235+164 \citep{Kutkin18}, indicating that the energy density of emitting particles dominates over that of magnetic field on parsec scales.

\subsection{Physical Parameters of Knot A}
\label{subsec:phpA}

The X-ray and radio spectral indices of knot~A differ from each other and equal $\alpha_{\text{R}}=1.32\pm0.17$ and $\alpha_{\text{X}}=0.66\pm0.15$, respectively \citep{Siem07}.
In the framework of a single electron energy distribution with the index $\gamma=2\alpha_{\text{R}}+1$, this distinction can be explained by the fact that the main contribution to the observed X-rays is given by ICS of part~2 of the CS spectrum with a restriction by the electron spectrum because $\alpha_{\text{X}}\approx\alpha_2$ within the errors  (Table~\ref{tab:input}).
Then, it follows that one of the electron distribution boundaries is such that the scattering on electrons with this boundary energy gives the main contribution to the observed X-ray flux. 
Since $2\alpha_2-\gamma+1<0$, the lower limit dominates in the final integration of Equation~(\ref{eq:Fcs_int}) over $d\Gamma$.
Thus, this boundary is lower and, as follows from Equation~(\ref{eq:osv}), $\Gamma_{\text{min}}$ lies within an interval from  $\Gamma_{\text{br}}=245$ to $\Gamma_0=1.5\cdot10^4$.

\begin{figure}
\begin{center}
\includegraphics[scale=0.45]{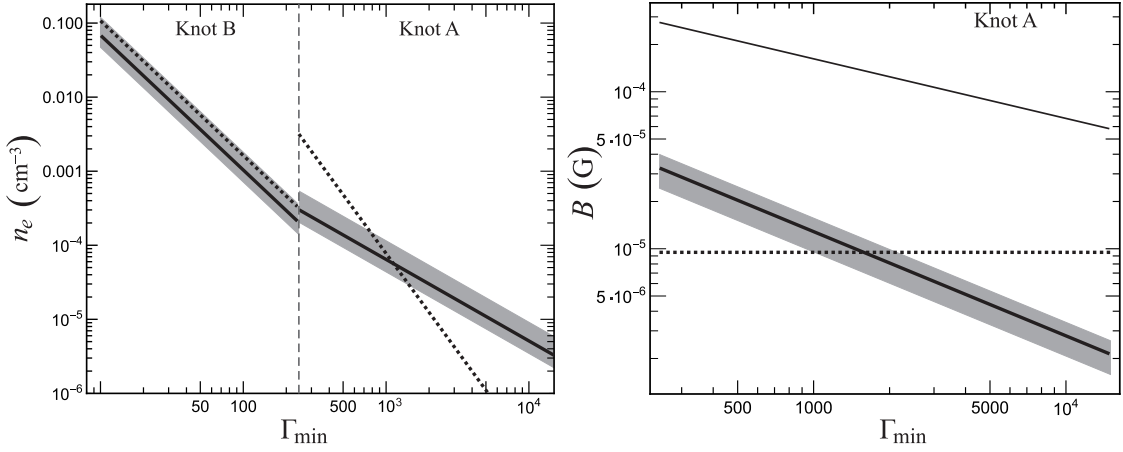}
\caption{Parameters of knot~A for the obtained interval of $\Gamma_{\text{min}}$ values. \textit{Left}: density of emitting electrons (right solid line). 
For comparison, we show the electron density in knot~B (left solid line) for its established interval of $\Gamma_{\text{min}}$ values. \textit{Right}: magnetic field for IC/CS (bold solid line) and equipartition magnetic field (thin solid line). Shaded areas show $n_e$ and $B$ values for $33^\circ<\theta<38^\circ$ (see Section~\ref{sec:kpcangle}). The electron density and magnetic field under IC/CMB are shown by dotted lines.
No correction for the Doppler factor of the kiloparsec-scale jet is made.
}
\label{fig:param}
\end{center}
\end{figure}

If $\Gamma_{\text{max}}<\Gamma_0$, then photons belonging only to part~2 of the CS spectrum are scattered to frequency $\omega_{\text{X}}$.
If $\Gamma_{\text{max}}>\Gamma_0$ (Fig.~\ref{fig:ph_e_distr}b), then electrons with $\Gamma_0<\Gamma<\Gamma_{\text{max}}$ scatter photons of part~1 of the CS spectrum to the observed X-ray frequency.
We compared X-ray fluxes originated by ICS of part~1 and 2 of the CS spectrum, $F_1^{\text{A}}$ and $F_2^{\text{A}}$, correspondingly. 
To find $F_1^{\text{A}}$ from Equation~(\ref{eq:Fcs_int}), we took into account that there is a restriction by the photon spectrum. 
To fulfill our case $F_2^{\text{A}}/F_1^{\text{A}}\gg1$, the value of $\Gamma_{\text{min}}\ll 7\cdot10^4$ is required, which corresponds to our conclusion about $\Gamma_{\text{min}}$ value and is always implemented for kiloparsec-scale jets.
Thus, the main contribution to the observed X-rays from knot~A is given by ICS of photons of part~2 of the CS spectrum on electrons with $\Gamma_{\text{min}}$ regardless of the value of $\Gamma_{\text{max}}$.
The expression for the expected IC/CS X-ray flux is similar to Equation~(\ref{eq:F2eB}) but with substitution of the corresponding knot volume $V$ and replacement of the projected distance $R_{\text{B}}$ on $R_{\text{A}}=2.87\cdot10^{23}$~cm, which corresponds to $11.2''$.   
Assuming, e.g., $\Gamma_{\text{min}}=10^3$, we obtain $n_e\approx6\cdot 10^{-5}$~cm$^{-3}$, magnetic field from the ratio of the radio to X-ray fluxes $B=1.3\cdot10^{-5}$~G, and equipartition magnetic field $B_{\text{eq}}=1.6\cdot10^{-4}$~G (see Figure~\ref{fig:param} and Table~\ref{tab:output}).

\section{Kiloparsec Jet Viewing Angle} \label{sec:kpcangle}

ICS of the central source radiation acting in knot~B implies that
\begin{equation}
F_{\text{IC/CS}}^{\text{B}}/F_{\text{IC/CMB}}>1. 
\label{eq:uneq1}
\end{equation}
The flux $F_{\text{IC/CS}}$ is defined by Equation~(\ref{eq:FxB}) and for convenient comparison with $F_{\text{IC/CMB}}$~(\ref{eq:Fcmb}) can be written in a similar form: 
\begin{equation}
\begin{split}
F_{\text{IC/CS}}^{\text{B}}=\frac{2^{(\gamma+1)/2}}{(\gamma+1)(2 \alpha_3+1-\gamma)}(1+z)^{-3-(\gamma-1)/2}\frac{3}{8 \pi} \sigma_{\text{T}} V \frac{c}{D_L^2}\left(m_e c^2 \right)^{1-\gamma} \mathcal{K} \times \\
\times \frac{L_{\text{CS}} \sin^2 \theta}{4 \pi c R^2_{\text{B}}} (1-\cos \theta)^{(\gamma+1)/2} \omega_{\text{br, j}}^{(\gamma-1)/2-1} \omega_{\text{X}}^{-(\gamma-1)/2},
\label{eq:FiccsB}
\end{split}
\end{equation}
where
\begin{equation}
L_{\text{CS}}=4\pi \left( 1+z\right)^{3+\alpha_3}\left(\frac{\delta_{\text{j}}}{\delta} \right)^{3+\alpha_3}\!\!\!D_L^2\,\, Q_3\,\, \omega_{\text{br, j}}^{-\alpha_3+1}=4.2\cdot 10^{49}~{\text{erg s}^{-1}} 
\label{eq:Lcs}
\end{equation}
is the CS luminosity at frequency $\omega_{\text{br, j}}$ in the kiloparsec-scale jet frame. 
Substituting Equations~(\ref{eq:FiccsB}) and~(\ref{eq:Fcmb}) into inequality~(\ref{eq:uneq1}), we found a lower limit for the interval of possible values of the PKS~1127$-$145 kiloparsec-scale jet viewing angle:
\begin{equation}
\theta_{\text{min}}^{\gamma+3}=\frac{2^{\gamma+1} (2\alpha_3+1-\gamma)}{ (\gamma+3)}W_{\text{CMB}} \frac{4 \pi c R_{\text{B}}^2}{L_{\text{CS}}} \left( \frac{\omega_{\text{CMB}}}{\omega_{\text{br, j}}} \right)^{(\gamma-1)/2-1}.
\label{eq:thetamin}
\end{equation}
Radio spectral indices $\alpha_{\text{R}}$ of knots~B and C are almost equal (see Table~\ref{tab:input}).
For each of these knots, $\alpha_{\text{R}}\approx\alpha_{\text{X}}$ within the measurement errors. Therefore we assume that IC/CS in knot~C occurs most efficiently on photons with frequency $\omega_{\text{br,\,j}}$ and use Equation~(\ref{eq:FxB}) with substitution of the corresponding $V$ and $R_{\text{C}}=7.3\cdot 10^{23}$~cm for the expression of the expected flux under IC/CS.
For knot~C, the pronounced X-ray brightness maximum is absent \citep{Siem07}.
Thus, we assume that X-ray emission of knot~C is generated by IC/CMB. 
Then the inequality, which is inverse to~(\ref{eq:uneq1}), is fulfilled and it defines an upper limit of the interval of possible $\theta$ values:
\begin{equation}
\theta_{\text{max}}^{\gamma+3}=\frac{2^{\gamma+1} (2\alpha_3+1-\gamma)}{(\gamma+3)}W_{\text{CMB}} \frac{4 \pi c R_{\text{C}}^2}{L_{\text{CS}}} \left( \frac{\omega_{\text{CMB}}}{\omega_{\text{br, j}}} \right)^{(\gamma-1)/2-1},
\label{eq:thetamax}
\end{equation}
where $\gamma=2.7$ is the spectral index of the electron energy distribution in knot~C (Table~\ref{tab:input}).

Thus, the kiloparsec-scale jet viewing angle is in the range $\theta_{\text{min}}=33^\circ<\theta<38^\circ=\theta_{\text{max}}$.
This is significantly different from $\theta<11^\circ$ obtained by \citet{Siem02} under the assumption of (i) straight jet and counter-jet, and (ii) apparent absence of the counter-jet due to relativistic beaming.
But, as the authors note, the kiloparsec-scale jet is bent, hence their angle estimation was crude, and $\theta>20^\circ$ may be valid for two distant knots~B and C.

Such a strong discrepancy between $\theta_{\text{pc}}=5^\circ$ and $\theta\approx35^\circ$ can be explained by the jet deceleration.
From the expression that connects angles in the parsec-scale jet frame (denoted by a prime) and in the observer's frame,
\begin{equation}
\sin \theta_{\text{pc}}'=\delta_{\text{pc}} \sin \theta_{\text{pc}}
\label{eq:angles}
\end{equation}
it follows that for the straight jet from parsec to kiloparsec scales $\theta'_{\text{pc}}=\theta'_{\text{kpc}}$, the Doppler factor of the kiloparsec-scale jet is
\begin{equation}
\delta_{\text{kpc}}=\delta_{\text{pc}}\frac{\sin \theta_{\text{pc}}}{\sin \theta_{\text{kpc}}}.
\label{dopkpc}
\end{equation} 
For $\delta_{\text{pc}}\approx11$, $\theta_{\text{pc}}=5^\circ$, $\theta_{\text{kpc}}=\theta\approx35^\circ$ we obtain $\delta_{\text{kpc}}\approx 1.7$, therefore, $\beta_{\text{kpc}}\approx0.8$ and $\Gamma_{\text{bulk}}\approx 1.7$. 
Hence, the jet on kiloparsec scales becomes moderately relativistic.

A change of the apparent jet viewing angle due to relativistic aberration occurs in the plane of the line of sight and jet velocity vector. 
A jet deceleration between parsec and kiloparsec scales cannot result in a change of the jet position angle.
As we noted in Sections~\ref{sec:iccspos} and~\ref{sec:iccs}, an initial angle of jet bend of $1^\circ-3^\circ$ can explain the observed difference in position angles for the parsec- and kiloparsec-scale jet. Then, the angle between the jets in the observer's frame is $\theta_{\text{pc,\,obs}}^{\text{kpc}}\approx 35^\circ$. The same angle in the reference frame of the kiloparsec-scale jet moving with speed $\beta_{\text{kpc}}$ is
\begin{equation}
\theta^{\prime\, \text{kpc}}_{\,\,\,\text{pc}}= \arctan\left[\frac{\beta_{\text{pc}}\sin \theta_{\text{pc}}^{\text{kpc}} \sqrt{1-\beta_{\text{kpc}}^2}}{\beta_{\text{pc}}\cos \theta_{\text{pc}}^{\text{kpc}}-\beta_{\text{kpc}}} \right]\approx 3^\circ .
\label{eq:SO}    
\end{equation}
In Section~\ref{sec:iccs} we assumed this angle to be equal to $1^\circ$. If the value of $\theta^{\prime\, \text{kpc}}_{\,\,\,\text{pc}}$ is three times greater, the magnetic field of knots~A and B decreases by $\lesssim10\%$ and the electron number density increases by $20-30\%$. Note that the estimation of electron number density is influenced mainly by the choice of the lower boundary of electron energy distribution, $\Gamma_{\text{min}}$ (see Fig.~\ref{fig:param}).
For $\theta^{\prime\, \text{kpc}}_{\,\,\,\text{pc}}=3^\circ$ the interval of the kiloparsec-scale jet viewing angle is $36^\circ-44^\circ$, which corresponds to $\delta_{\text{kpc}}\approx1.5$, $\beta_{\text{kpc}}\approx0.7-0.8$ and $\Gamma_{\text{kpc}}\approx1.5$.
A more precise determination of the geometric and physical parameters of the kiloparsec-scale jet seems impractical due to errors in the observed parameters and the presence of two solutions for  $\beta_{\text{kpc}}$ from knowing $\delta_{\text{kpc}}$ and $\theta_{\text{kpc}}$.
Thus, our estimation of the physical parameters of knots~A and B under IC/CS and spatial orientation of the kiloparsec-scale jet is self-consistent.

\section{Prediction on the Gamma-Ray Flux}
\label{sec:gam}

As noted by \citet{Georg06}, gamma-ray observations of quasars can be used to test high-frequency emission models of their jets. Therefore, in this section we model an expected high-frequency spectrum of jet knots of the quasar PKS~1127$-$145 under IC/CS to compare with the jet total gamma-ray flux measured by \textit{Fermi}-LAT. 

\begin{figure}
\begin{center}
\includegraphics[scale=0.5]{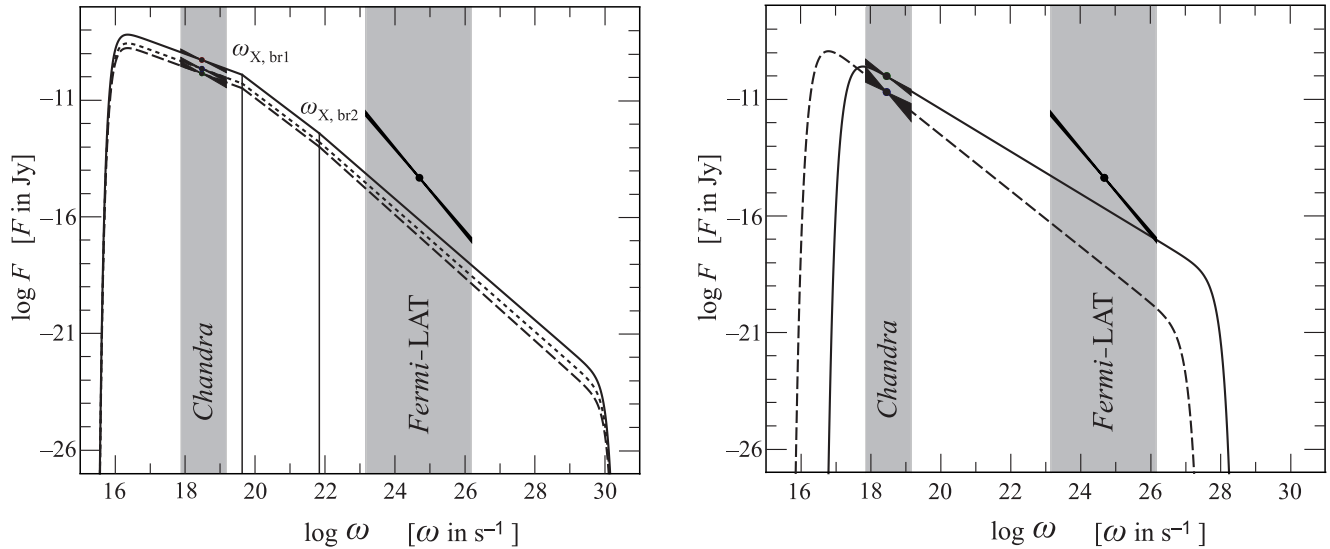}
\caption{
High-energy emission spectrum of the jet knots of PKS~1127$-$145 under IC/CS model except the most distant knot~C for which IC/CMB was applied.  Left panel: spectra for knots~I (solid line), O (dashed line), and A (dotted line). Two spectral breaks marked by vertical lines for each of these knots are caused by (i) the difference in contribution from various parts of the CS spectrum to scattered flux at different frequencies and (ii) the transition from a restriction set by the minimum Lorentz factor of the electron energy distribution to one set by the upper limit of the scattering spectrum. Right panel: spectra for knots~B (solid line) and C (dashed line). On both panels shaded gray areas correspond to the \textit{Chandra} and \textit{Fermi}-LAT ranges. ``Bow-ties'' in the \textit{Chandra} range show the observed X-ray flux of the knots.
The ``bow-tie'' in the \textit{Fermi}-LAT range illustrates the spectral flux corresponding to the historical minimum of the gamma-ray flux of PKS~1127$-$145 \citep{3FGL}, which we assume as an upper limit to the gamma-ray emission of the kiloparsec-scale jet, and corresponds to the spectral index $1.79\pm0.05$.
} % caption
\end{center}
\label{fig:predict_gamma}
\end{figure}

The X-ray spectral indices $\alpha_{\text{X}}$ of the innermost knots~I, O, A are similar, namely $0.67\pm0.11$, $0.69\pm0.19$, and $0.66\pm0.15$ \citep{Siem07} and agree within the errors with the spectral index $\alpha_2=0.55\pm0.13$ of part~2 of the CS spectrum (Table~\ref{tab:input}), while the radio spectral indices $\alpha_{\text{R}}$ of knots~A ($\alpha_{\text{R}}=1.32$) and O ($\alpha_{\text{R}}>1.35$) differ from $\alpha_{\text{X}}$. For knot~I, $\alpha_{\text{R}}$ is not measured due to a lack of angular resolution to distinguish it from the bright core \citep{Siem02,Siem07}. 
We assume $\alpha_{\text{R}}\gtrsim1.3$ because the typical spectral index of bright jet features on parsec scales is about 0.7~\citep{Pushkarev12} and it becomes steeper with distance from the CS due to energy losses \citep{Kardashev62}, the effect known as spectral aging. 
Thus, as follows from Equations~(\ref{eq:wj}) and~(\ref{eq:osv}), the scattering of part~2 of the CS spectrum on electrons with $\Gamma_{\text{min}}$ dominates up to a frequency $\omega_{\text{X,br1}}=(4/3)(\delta_{\text{j}}/\delta)\omega_{\text{br}}\Gamma_{\text{min}}^2$, that for, e.g., $\Gamma_{\text{min}}=10^3$ is $4.3\cdot10^{19}$~s$^{-1}$, which is slightly higher than the upper limit of the \textit{Chandra} frequency range.
Emission at frequencies higher than $\omega_{\text{X, br1}}$ arises from both the scattering of part~2 of the CS spectrum with the photon spectrum restriction and the scattering of part~3 of the CS spectrum. 
In the latter case, according to Equation~(\ref{eq:osv}), the restriction by the electron spectrum acts at frequencies lower than $\omega_{\text{X,\,br2}}=4/3\omega_{\text{max,\,j}}\Gamma_{\text{min}}^2(1+z)^{-1}$. Thus, ICS on electrons having $\Gamma_{\text{min}}$ contributes mainly to the scattered emission, with spectral index  $\alpha_3$. At frequencies above $\omega_{\text{X,\,br2}}$ the emission is produced by ICS of photons with frequency $\omega_{\text{max,  j}} =1.1\cdot10^{16}$~s$^{-1}$ on electrons with Lorentz factor greater than $\Gamma_{\text{min}}$. 
The spectral index of this emission is $(\gamma-1)/2=\alpha_{\text{R}}$. Comparison of the scattered fluxes shows that at $\omega>\omega_{\text{X,\,br1}}$, the flux produced by ICS of photons of part~2 of the CS spectrum is $\lesssim 10\%$ of that formed by scattering of photons of part~3 of the CS spectrum. Figure~\ref{fig:predict_gamma} (left panel) shows the expected spectra of knots~I, O, A with the following parameters: $\Gamma_{\text{min}}=10^3$, $\Gamma_{\text{max}}=10^7$, $\omega_{\text{br, j}}=7\cdot10^{13}$~s$^{-1}$.
Note that the value of $\Gamma_{\text{min}}$ influences the frequencies of the spectral breaks of the knots. The range of possible values of $\Gamma_{\text{min}}$ (see Sect.~\ref{subsec:phpA}) is such that the breaks occur at frequencies lower than those covered by the \textit{Fermi}-LAT range.

For knot~B, the main contribution to the high-frequency emission is brought about by ICS of photons with a frequency that corresponds to the high-energy break of the CS spectrum (see Sect.~\ref{subsec:phpB}).
Almost the same case is the knot~C, though the high-frequency emission from this knot is originated mostly by ICS of photons with the frequency of the spectral maximum of the CMB at a given redshift. The spectra of knots~B and C shown in Figure~\ref{fig:predict_gamma} (right panel) were constructed using $\Gamma_{\text{min}}=100$, $\Gamma_{\text{max}}=10^7$, $\omega_{\text{br, j}}=7\cdot10^{13}$~s$^{-1}$, and $\omega_{\text{CMB}}=2.2\cdot10^{12}$~s$^{-1}$.
For all spectra, the parameter $\Gamma_{\text{max}}$ defines the frequency of the high-energy cutoff only. 
The gamma-ray flux integrated over the \textit{Fermi}-LAT frequency range from all knots of the jet is $8.1\cdot10^{-2}$~eV\, cm$^{-2}$\,s$^{-1}$, while the same flux from knot~B is $5.1\cdot10^{-2}$~eV\,cm$^{-2}$\,s$^{-1}$.

Further, we compare this estimated gamma-ray total jet flux with the observed data obtained by \textit{Fermi}-LAT. If the expected jet flux is smaller than or equal to the observed flux from the jet, our model of kiloparsec-scale jet X-ray emission has no contradiction in gamma-rays. In the gamma-ray range the angular resolution is too low to resolve the kiloparsec-scale jet from the core. An upper limit to the flux from the kiloparsec-scale jet can be obtained by selecting the constant flux level from observed light curve \citep{Georg06, MeyGeor14, Meyer15}. As PKS~1127$-$145 shows significant variations in the gamma-ray flux, we use the minimum value as the upper limit on the flux from the kiloparsec-scale jet because there is no reliable way to estimate what portion of this value is produced by the kiloparsec-scale jet emission.
To derive the minimum energy flux of the source we made use of the 3FGL data \citep{3FGL}, namely 48 measurements of monthly binned photon flux integrated from 100~MeV to 100~GeV and photon index $\Gamma_{\text{ph}}=2.79\pm 0.05$. The minimum gamma-ray energy flux of PKS~1127$-$145 estimated this way is about 3~eV\,cm$^{-2}$\,s$^{-1}$, which is significantly higher than the total predicted flux level of the kiloparsec-scale jet. 
Therefore, IC/CS cannot be ruled out by the gamma-ray observational data.

\section{Discussion}
\label{sec:disc}

\citet{Cel01} assumed IC/CS as a possible X-ray emission mechanism for kiloparsec-scale jets of radio galaxies. 
For this AGN class, two-sided radio jets are observed. 
This indicates a rather large jet angle to the line of sight to dominate energy density of the CS emission over energy density of the CMB.
In the radio band mainly one-sided kiloparsec-scale jets are observed for core-dominated quasars. 
This is interpreted in terms of the Doppler de-boosting of the counter-jet \citep{Tav00}.
In this case, small angles between kiloparsec-scale jets and the line of sight are expected. 
For the observed projected distance from the core, this implies such a great distance from the central source that its energy density is less than that of the CMB. 
Therefore, the assumption about the origin of X-rays as a result of IC/CMB was reasonable and explained the shape of yhe spectral energy distribution for most kiloparsec-scale jet knots of core-dominated quasars \citep{Tav00, Cel01}.

However, difficulties arise in interpretation of the observational data.   
First, the observed X-ray flux is produced by electrons with a very small Lorentz factor ($\sim 1-10$) \citep{Tav00, Harris06}.
Synchrotron emission of these electrons has a low frequency which is not accessible for observations.
Therefore, to estimate the physical parameters of the emission region, the ultra-relativistic electron spectrum is assumed to retain a power-law.
But if it becomes flatter or even has a low-energy cutoff, then a smaller jet viewing angle and higher bulk jet velocity are required \citep[see][and references therein]{Harris06}. 
Second, as $W_{\text{CMB}}$ increases with redshift, one would expect a higher density of AGN jets detected in X-rays at higher $z$, but this is not confirmed by observations \citep{KatSta05}.
Third, a decrease in X-ray intensity with distance from the core observed in jets is interpreted by the common deceleration of a jet \citep{GeorKaz04}.   
In this context, no obvious reasons for the existence of jet knots are present \citep{Harris06}.
Furthermore, the beamed IC/CMB model implies that there are no bending and deceleration of a jet from parsec to kiloparsec scales.      
Hence, if a kiloparsec-scale jet of a core-dominated quasar has no detected X-rays then it has bending and/or deceleration from parsec to kiloparsec scales. 
This would be found indirectly in the distinction of parsec-to-kiloparsec jet position angle distributions, $\Delta\text{PA}$, for quasars with and without the detected X-ray jets.
However, no statistically significant difference for $\Delta\text{PA}$ distributions has been detected \citep{But16}.

The beamed IC/CMB model might finally be excluded for 3C~273 \citep{MeyGeor14} and PKS~0637$-$752 \citep{Meyer15} because no high-level gamma-ray flux, expected within this model, is detected.    
Although the expected gamma-ray flux can be reduced by taking into account electron cooling \citep{LTC17}, optical polarization observations set tight constraints. 
Namely, high optical polarization in jet knots of PKS~1136$-$135 can be explained within IC/CMB if $\Gamma_{\text{min}}\sim 1$ and $\delta_{\text{kpc}} \ge 20$ \citep{Cara13}.   

We assume that the X-ray emission mechanism is ICS of the CS photons in at least two inner jet knots of PKS~1127$-$145, and the CS emission from the radio to millimeter ranges comes from the parsec-scale jet.
A natural decrease of the scattering emission flux with distance and insensitivity of ICS to the magnetic field explain the observed radio and X-ray intensity distributions along the kiloparsec-scale jet.  
Spatial coincidence of the radio and X-ray locations of the corresponding jet knots \citep{Siem07} is caused by the fact that electrons, which belong to the same energy distribution, produce synchrotron radio emission and X-rays through ICS. 
This is true even for knot~A, which has different radio and X-ray spectral indices whereas, to explain the spectral index distinction within the beamed IC/CMB model, \citet{Samb04} assume the power-law electron spectrum to become flatter near the low-energy cutoff.  
Synchrotron X-ray models introduce a second high-energy electron population that reflects the action of the second acceleration mechanism \citep[see, e.g.][]{Jester06}.
Within this, as \citet{Schwartz2000} note, no physical reasons for the spatial coincidence of two independent particle distributions are present for the observed knots.

The IC/CS and IC/CMB scenarios competing within the same jet allow us to derive the kiloparsec-scale jet viewing angle $\theta$.
We note that the angle estimated this way is independent from the bulk kiloparsec-scale jet velocity. 
Thus, for the PKS~1127$-$145 jet, we obtain $\theta \approx 35^\circ$ and $\beta_{\text{kpc}}\approx0.8$ ($\delta\approx1.7$).
\citet{MBK10} made a similar estimation of the kiloparsec-scale jet viewing angle for the quasar 3C~273 and found $\theta\approx30^\circ$.
This significantly differs from the value $\theta\approx5^\circ$ for $\delta_{\text{kpc}}\approx5$ \citep{Harris06} obtained within the beamed IC/CMB model.
But for the kiloparsec-scale jet of the quasar 3C~273, \citet{CD94} found $\theta=30-35^\circ$ and $\beta_{\text{kpc}}=0.8$ based on photometric and polarimetric observational data and applying magnetic hydrodynamic equations. 
 \citet{FC85} have shown that the interpretation of high radio polarization of the terminal 3C~273 jet knot requires $\theta\sim45^\circ$. 
Hence, for PKS~1127$-$145, the resulting estimation of $\theta\approx35^\circ$ is adequate.  
The difference between $\theta_{\text{pc}}$ and $\theta$ can be explained by the fact that, in the comoving frame of the relativistic moving object, the angle between the velocity vector and the line of sight is larger than the same angle in the observer's frame; the latter increases as the object velocity decreases.
This change of the angle occurs in the plane containing the line of sight.
Then, an actual jet bend of $1^\circ-3^\circ$ between parsec and kiloparsec scales in the observer's frame and deceleration of the kiloparsec-scale jet down to the bulk velocity $\beta_{\text{kpc}}\approx0.8$ are enough to interpret both the distinction of $\theta_{\text{pc}}$ with $\theta$ and the observed different position angles for the jet on parsec and kiloparsec scales.
Analysis of jet and counter-jet for 13 active galaxies reveals that jet speeds are in the range from 0.6$c$ to 0.7$c$ and its upper limit $<0.95c$ ($\Gamma\leq 3$) \citep{WA97}.
Moderate relativistic speeds of kiloparsec-scale jets $\gtrsim0.6c$ and $\lesssim0.75$ were derived by \citet{ArLon04} and \citet{MullinH09}.
Furthermore, according to VLBA observations, the deceleration is registered starting from distances of about 100~pc \citep{Homan15}.
Note that we assume in the terminal knot~C that X-rays are produced by IC/CMB. 
If IC/CS gives the main contribution to X-rays of this knot, then the jet viewing angle increases, resulting in a decrease of the knot~C separation from the CS.
This leads to an increase of $B$ and decrease of $\delta_{\text{kpc}}$ and $n_e$. 
Thus, the electron density and magnetic field strength approach to their equipartitional values. 

Taking into account the characteristic of scattering radiation and assuming IC/CS, we found the magnetic field strength $B$ and density of radiating electrons $n_e$ in A and B~jet knots of PKS~1127$-$145.
These values are of the same order of magnitude as those obtained within IC/CMB for $\delta_{\text{kpc}}=1$ and they correspond to the case when the energy density of particles is higher than that of the magnetic field. 
\citet{Siem02} note that assuming electrons only in the jet, $\delta_{\text{kpc}}\approx2$ is required to fulfill the equipartition. 
For $\delta_{\text{kpc}}=1.7$ obtained by us, the knot parameters of the PKS~1127$-$145 jet become closer to the equipartition values, but the particle energy density dominates.
Following \citet{Pachol}, for simplicity we ignored the angle between the magnetic field direction and the line of sight.
However, taking into account this angle results in even stronger domination of particle energy (see Eq.~(\ref{eq:Heq})). 
We suppose that there is no need to achieve energy density equipartition because, first, the assumption about the energy equipartition was introduced by \citet{Burb} to separate $B$ and $n_e$ and to derive estimations of these values using the observed synchrotron emission.
The author emphasize that equipartition can be unrealizable for powerful kiloparsec-scale jets.
Second, the results of recent observations carried out by the Earth-space radio interferometer \textit{RadioAstron} for several sources show extreme brightness temperatures that exceed the limit established by IC cooling under equipartition \citep{Gomez16, Kovalev16, Pilipenko18, Kutkin18}.   
It follows that either equipartition is not present or the Doppler factor for parsec-scale jets is $\delta\sim100$, which is unlikely.

An additional advantage of IC/CS over IC/CMB is as follows.
IC/CS predicts a level of the steady gamma-ray flux that agrees with observational data. 
There is no strict dependence on redshift for parameters of the CS spectrum such as frequency of the peak flux or frequency of the radio-to-millimeter spectral break.   
Hence, within IC/CS there is no contradiction between the expected and observed number of X-ray jets in dependence on $z$.
\citet{Cara13} observed high optical polarization in knots of the PKS~1136$-$135 jet and found the direction of the electric vector to differ from that obtained from the radio data. 
This can be due to some contribution made by IC/CS into the optical emission from these knots.
This polarization issue needs a detailed study, for example, according to the formalism of Nagirner \citep{NagirnerP94,NagirnerP01,Nagirner94} but it is beyond the scope of this paper.

For core-dominated quasars, the X-ray intensity decreases, while the radio intensity increases with distance from the CS \citep{Harris06, Samb04}.
Additionally, the X-ray spectrum is flatter than the radio spectrum for knots closer to the core.
For interpretation of these facts, IC/CS can be a single universal mechanism that does not require additional assumptions.
Furthermore, IC/CS and competition of IC/CS and IC/CMB that takes place in a single jet can be a useful tool for determination of the jet viewing angle and jet velocity on kiloparsec scales if there are observational data on (i) knots in the radio and X-ray bands, (ii) the radio-to-millimeter spectrum of the central source.

\section{Conclusions}
\label{sec:conc}
		
We have investigated ICS of CS photons as a possible X-ray emission mechanism for the quasar PKS~1127$-$145 jet.
IC/CS gives the simplest explanation for both the observed intensity distributions along the jet and spectral indices of knots in the radio and X-ray domains within an assumption of a single power-law electron energy distribution. 
According to this scenario, when scattering on electrons with a fixed energy gives the main contribution to the scattered flux, the spectral indices of radio and X-ray emission are different, with  the latter equal to the spectral index of the scattered emission. 
When photons with a fixed frequency are most efficiently scattered, the spectral indices of radio and X-ray radiation are equal and determined by the spectral index of the power-law electron energy distribution.

We found an interval of a possible value for the kiloparsec-scale jet angle of PKS~1127$-$145 with the line of sight to be $33^\circ\lesssim\theta\lesssim38^\circ$. 
IC/CMB is supposed to dominate in the most distant knot C.
To match the kiloparsec-scale jet viewing angle with both the parsec-scale jet viewing angle and the observed position angles of the jet on parsec and kiloparsec scales, we determine the jet bend angle between these scales to be $1^\circ-3^\circ$ and the kiloparsec-scale jet to have a bulk velocity of about $0.8c$.

IC/CS provides reasonable estimates for the magnetic field and density of emitting electrons and agrees with all available observational data including the prediction of the adequate flux level from jet knots in the gamma-ray range.   
The most realistic way to test IC/CS is a search for the X-ray spectral break, similar to that for knot~A of the PKS~1127$-$145 jet, for knots of other core-dominated quasar jets.
Another independent confirmation of IC/CS is the good agreement between the theoretically expected polarization under IC/CS with that observed in the optical range assuming that at these frequencies the ``tail" of high-energy emission gives the main contribution to the detected flux.

\acknowledgements

M.S.B. is grateful to Victor M. Kontorovich for inspiration for this research. 
We thank the anonymous referee for useful comments, which helped to improve the
manuscript. This research is supported by the Russian Foundation for Basic Research, project No.~18-32-00824. Under this project the following results have been obtained: (i)  IC/CS as an X-ray emission mechanism was supposed; (ii) formulae for the spectral flux of scattered emission for IC of both power-law photon  and electron energy distributions were derived from the Boltzmann transport equation; (iii) the difference of the spectral indices of radio and X-ray emission of knot~A was explained under a single power-law electron energy distribution; (iv) electron number density, magnetic field, expected gamma-ray flux and spectrum were estimated; (v) viewing angle and speed of the kiloparsec-scale jet were found. Analysis of the \textit{Fermi}-LAT data was supported by the Academy of Finland projects 296010 and 318431.

\vspace{5mm}
\facilities{{\it Fermi}-LAT}  

\appendix

\section{Boltzmann Transport Equation for Inverse Compton Scattering} 
\label{sec:app}

We use the Boltzmann transport equation for ICS in the invariant form written by Nagirner \citep{NagirnerP94,NagirnerP01,Nagirner94}.
Neglecting the decrease of soft photon number due to scattering and the induced scattering, for the stationary case we obtain:
\begin{equation}
 \mbox{\boldmath $k'$} \nabla N(\mbox{\boldmath $k'$})=\int \frac{r_e^2}{2} (m_e c^2)^2 \frac{\hbar c^2}{E E' \omega} \delta (\mbox{\boldmath $p$}+\mbox{\boldmath $k$}-\mbox{\boldmath $p'$}-
 \mbox{\boldmath $k'$}) \delta (E+\hbar \omega-E'-\hbar \omega' ) G(\xi,\xi')f(\mbox{\boldmath $p$}) N (\mbox{\boldmath $k$}) d^3 \mbox{\boldmath $p'$}\, d^3  \mbox{\boldmath $p$}\, d^3 \mbox{\boldmath $ k $},
\label{eq:kineq1}
\end{equation}
where the values after scattering are denoted by prime, \mbox{\boldmath $p$}, \mbox{\boldmath $k$} and $E$, $\hbar \omega$ are the momenta and energies of electron and photons before scattering, respectively, $r_e$ and $m_e$ are the classical radius and the mass of electron, $c$ is the speed of light, $f(\mbox{\boldmath $p$})$, $N (\mbox{\boldmath $k$})$ are the distribution functions of electrons and photons, respectively, and 
\begin{equation}
\begin{split}
G(\xi,\xi')&=(1/\xi-1/\xi')^2+2(1/\xi-1/\xi')+\xi/\xi'+\xi'/\xi, \\
\xi&=\frac{E \hbar \omega}{m_e^2 c^4} (1-\beta \cos \psi), \,\,\,\, \xi'=\frac{E \hbar \omega'}{m_e^2 c^4} (1-\beta \cos \phi),
\label{eq:Gxi}
\end{split}
\end{equation}
where $\beta~\approx~1$ is the speed of the ultra-relativistic electrons in units of $c$, and $\psi$ and $\phi$ are the angles between electron and photon momenta before and after
scattering, respectively. Under ICS, $\phi$ is small, $\psi\approx\theta$, where $\theta$ is the angle between photon momenta before and after scattering.
Since we are interested in photons scattered to an observer, after integration over $d^3 \mbox{\boldmath $ p'$}$ using first a $\delta$-function and taking into account a small change in electron energy under one act of scattering, Equation~(\ref{eq:kineq1}) becomes 
\begin{equation}
\frac{d}{d s}N (\mbox{\boldmath $k'$})=\int \frac{r_e^2}{2} (m_e c^2)^2 \frac{c^3}{E^2 \omega \omega'^2} \delta((1-\beta \cos \phi)-\frac{\omega}{\omega'}(1-\beta \cos \psi)) G(\xi,\xi')f(\mbox{\boldmath $p$}) N (\mbox{\boldmath $k$}) d^3  \mbox{\boldmath $ p$}\, d^3 \mbox{\boldmath $ k $},
\label{eq:kineq2}
\end{equation}
where $s$ is the length of the radiating region along the line of sight.

The spectral flux scattered to the observer's direction is equal to
\begin{equation}
F(\omega')=c \hbar \omega' N(\omega') \frac{x y}{D^2_L}=\frac{\hbar \omega'^3}{c^2} \frac{x y}{D^2_L}N(\mbox{\boldmath $k'$}),
\label{eq:Fws}
\end{equation}
where $x$, $y$ are the observed knot sizes, and $D_L$ is the quasar luminosity distance. 

From the transport Equation~(\ref{eq:kineq2}) we obtain an expression for the expected X-ray flux under IC/CMB. The CMB spectrum can be considered to be monochromatic. Then 
\begin{equation}
N(\mbox{\boldmath $k$})=\frac{c^3}{4 \pi \omega^2} N(\omega),\,\,\,N(\omega)=\frac{W_{\text{CMB}}}{\omega_{\text{CMB}}}\, \delta(\omega-\omega_{\text{CMB}}),
\label{eq:Ncmb}
\end{equation}
where $\omega_{\text{CMB}}$ is the frequency of the peak in the microwave background spectrum at the object's redshift.

In a simple case, we assume that the knot electron distribution is uniform and isotropic. Then
\begin{equation}
f(\mbox{\boldmath $p$})=\frac{c^3}{4 \pi E^2 } f(E), \,\,\,\,f(E)=\mathcal{K}E^{-\gamma}.
\label{eq:fe}
\end{equation}
Substituting Equations~(\ref{eq:Ncmb}) and (\ref{eq:fe}) into Equation~(\ref{eq:Fws}) and integrating over $d^3 \mbox{\boldmath $p$}=(E^2/c^3) 2 \pi dE d (1-\beta \cos \phi)$ and $d^3 \mbox{\boldmath $k$}= (\omega^2/c^3) 2 \pi d\omega d(1-\beta \cos \psi)$ in the interval for $\phi$ and $\psi$ from 0 to $\pi$, for the observed flux we obtain
\begin{equation}
F_{\text{IC/CMB}}(\omega_{\text{X}})=\frac{2^{\gamma+1}}{(\gamma+1)(\gamma+3)} (1+z)^{-3-(\gamma-1)/2} \frac{3}{8 \pi}\, c\, \sigma_{\text{T}} \frac{V}{D_L^2}(m_e c^2)^{1-\gamma} \mathcal{K}W_{\text{CMB}}\, \omega_{\text{CMB}}^{(\gamma-1)/2-1}\omega_{\text{X}}^{-(\gamma-1)/2},
\label{eq:Fcmb}
\end{equation}
where $\sigma_{\text{T}}=(8\pi/3)r_e^2$ is the Thomson cross section, $V=xy^2/\sin \theta$ is the volume of the emitting region, $\theta$ is the jet viewing angle, $W_{\text{CMB}}$ is the energy density of the cosmic microwave background, $\omega_{\text{X}}$ is the observed X-ray frequency.  
Formula~(\ref{eq:Fcmb}) up to the numerical factor coincides with the expression obtained from the average electron radiating losses under ICS \citep[see, e.g., the Equation~(9) in][]{MBK10}.

\bibliography{pks1127m145}

\end{document}